\documentstyle[preprint,prl,aps]{revtex}

\def\Cmae{\mbox{\hspace*{-10pt}}}
\def\Usr{\mbox{\hspace*{20pt}}}
\def\Cusr{\mbox{\hspace*{10pt}}}
\def\Define{\mathop{\stackrel{\rm def}{=}}}
\newcommand{\myref}[1]{(\ref{#1})}
\newcommand{\vecvar}[1]{\mbox{\boldmath$#1$}}
\begin{document}
\draft
\title{New Non-Symmetric Orthogonal Basis for the Calogero Model\\
with Distinguishable Particles}
\author{Hideaki Ujino~\cite{e-mail}, Akinori Nishino and Miki Wadati}
\address{Department of Physics, Graduate School of Science,
University of Tokyo,\\
Hongo 7--3--1, Bunkyo-ku, Tokyo 113--0033, Japan}
\date{\hspace*{7cm}}
\maketitle
\begin{abstract}
We demonstrate an algebraic construction of all the simultaneous
eigenfunctions of the conserved operators for distinguishable particles
governed by the Calogero Hamiltonian. Our construction is completely parallel
to the construction of the Fock space for decoupled quantum harmonic
oscillators. The simultaneous eigenfunction does not coincide with
the non-symmetric Hi-Jack polynomial, which shows that the conserved
operators derived from the number operators of the decoupled quantum
harmonic oscillators are algebraically different from the known ones derived
by the Dunkl operator formulation.
\end{abstract}

There has been a surge of interest in 
the orthogonal symmetric polynomials associated with the quantum
integrable systems with inverse-square long-range interactions.
Particularly after the Jack polynomials, which span the orthogonal basis
for the Sutherland model~\cite{Sutherland_1,Jack_1,Stanley_1,Macdonald_1},
enabled an exact calculation of dynamical density-density correlation
functions of the model~\cite{Ha_1}, the Jack polynomial and
its variants have been extensively studied.
So far, we have been studying
the Hi-Jack (or multivariable Hermite)
symmetric polynomial that forms the symmetric orthogonal basis
for the Calogero model~\cite{Calogero_1,Ujino_4,Ujino_1,Ujino_2,Ujino_3},
\begin{equation}
  \hat{H}_{\rm C}=\frac{1}{2}\sum_{j=1}^{N}
  \bigl(p_{j}^{2}+\omega^{2}x_{j}^{2}\bigr)
  +\frac{1}{2}\sum_{\stackrel{\scriptstyle j,k=1}{j\neq k}}^{N}
  \frac{a^{2}-aK_{jk}}{(x_{j}-x_{k})^{2}},
  \label{eqn:Calogero_Hamiltonian}
\end{equation}
where $p_{j}=-{\rm i}\frac{\partial}{\partial x_{j}}$.
The operator $K_{jk}$ is a coordinate exchange operator that is
defined by the action on a multivariable function,
$(K_{jk}f)(\cdots,x_{j},\cdots,x_{k},\cdots)
=f(\cdots,x_{k},\cdots,x_{j},\cdots)$.
For the symmetric case, the exchange operators at the rightmost
of the expressions are identified with
the identity, $K_{jk}\equiv 1$.
Using the Dunkl operator formulation~\cite{Dunkl_1,Polychronakos_1},
we derived the Rodrigues formula for the Hi-Jack symmetric
polynomial for the first time~\cite{Ujino_1}.
The Hi-Jack symmetric polynomials are
the simultaneous eigenfunctions of all the commuting conserved operators
of the Calogero model derived by the Dunkl operator
formulation~\cite{Ujino_2,Ujino_3}.
They describe the indistinguishable particles that
obey the Calogero Hamiltonian and are considered to be multivariable
Hermite polynomials with one additional parameter~\cite{Ujino_2}.
The name Hi-Jack comes from the fact that the symmetric polynomial is
a one-parameter deformation of the Jack polynomial.
Sogo showed a transformation of the symmetric Calogero
Hamiltonian into the Euler operator
${\cal O}_{\rm E}\Define\sum_{j=1}^{N}x_{j}\frac{\partial}{\partial x_{j}}$%
~\cite{Sogo_1}.
Recently, Gurappa and Panigrahi showed a transformation of the
symmetric Calogero
Hamiltonian into decoupled quantum harmonic oscillators~\cite{Gurappa_1}.
Motivated by their results, we presented an algebraic construction of a
new symmetric orthogonal basis for the Calogero Hamiltonian~\cite{Ujino_5}
in a completely parallel way to the construction of the Fock space of
the bosonic quantum harmonic oscillators.
Since the first quantized Hamiltonian itself has no information on the
statistics of the particles, there should be eigenfunctions not only for
indistinguishable particles but also for distinguishable particles.
Namely, there should be non-symmetric eigenfunctions that describe
distinguishable particles governed by the Calogero Hamiltonian
\myref{eqn:Calogero_Hamiltonian}.
In this letter, we shall show a simple algebraic formula for all the
simultaneous eigenfunctions of the conserved operators of the
Calogero model with distinguishable
particles and discuss the meaning of the transformation into
the interaction-free system.

First, we decompose the Calogero Hamiltonian
\myref{eqn:Calogero_Hamiltonian} into interaction-free $N$-particle
Hamiltonian.
The ground state wave function of the Calogero model
is given by the real Laughlin wave function,
\[
  \phi_{\rm g}(\vecvar{x})=\prod_{1\leq j<k\leq N}|x_{j}-x_{k}|^{a}
  \exp\Bigl(-\frac{1}{2}\omega\sum_{j=1}^{N}x_{j}^{2}\Bigr),
\]
and the ground state energy is given by
$E_{\rm g} = \frac{1}{2}N\omega\bigl(Na+(1-a)\bigr)$.
Then the following similarity transformation removes the action to the ground
state wave function from the Hamiltonian,
\[
  \phi_{\rm g}^{-1}(\hat{H}_{\rm C}-E_{\rm g})\phi_{\rm g}
  = \omega{\cal O}_{\rm E}-\frac{1}{2}{\cal O}_{\rm L},
\]
where the non-symmetric Lassalle operator
${\cal O}_{\rm L}$ is~\cite{Baker_2}
\begin{eqnarray*}
  {\cal O}_{\rm L} & = & 
  \sum_{j=1}^{N}\frac{\partial^{2}}{\partial x_{j}^{2}}
  +a\sum_{\stackrel{\scriptstyle j,k=1}{j\neq k}}^{N}
  \Bigl(\frac{1}{x_{j}-x_{k}}
  \bigl(\frac{\partial}{\partial x_{j}}-\frac{\partial}{\partial x_{k}}
  \bigr)\nonumber\\
  & &+\frac{K_{jk}-1}{(x_{j}-x_{k})^{2}}\Bigr).
\end{eqnarray*}
As is similar to Sogo's approach to the symmetric case,
the Calogero Hamiltonian can be transformed to the Euler operator.
The commutation relation
$\bigl[{\cal O}_{\rm L},{\cal O}_{\rm E}\bigr]=2{\cal O}_{\rm L}$
and the Baker-Hausdorff formula yield
\[
  {\rm e}^{\frac{1}{4\omega}{\cal O}_{\rm L}}
  \bigl(\omega{\cal O}_{\rm E}-\frac{1}{2}{\cal O}_{\rm L}\bigr)
  {\rm e}^{-\frac{1}{4\omega}{\cal O}_{\rm L}}=\omega{\cal O}_{\rm E}.
\]
In a similar way, we can transform the Euler operator into the Hamiltonian
of the decoupled quantum harmonic oscillators.
The commutator between the Euler operator ${\cal O}_{\rm E}$ and
the Laplacian
$\nabla^{2}\Define\sum_{j=1}^{N}\frac{\partial^{2}}{\partial x_{j}^{2}}$,
$\bigl[\nabla^{2},{\cal O}_{\rm E}\bigr]=2\nabla^{2}$,
and again the Baker-Hausdorff formula yield
\[
  {\rm e}^{-\frac{1}{4\omega}\nabla^{2}}
  \omega{\cal O}_{\rm E}
  {\rm e}^{\frac{1}{4\omega}\nabla^{2}}=
  \omega\bigl({\cal O}_{\rm E}-\frac{1}{2\omega}\nabla^{2}\bigr).
\]
Another similarity transformation using the Gaussian kernel produces
the Hamiltonian for the decoupled $N$ quantum harmonic oscillators,
\[
  {\rm e}^{-\frac{1}{2}\omega\vecvar{x}^{2}}
  \omega\bigl({\cal O}_{\rm E}-\frac{1}{2\omega}\nabla^{2}\bigr)
  {\rm e}^{\frac{1}{2}\omega\vecvar{x}^{2}}
  = \frac{1}{2}\Bigl(
  \sum_{j=1}^{N}\bigl(p_{j}^{2}+\omega^{2}x_{j}^{2}\bigr)
  -N\omega\Bigr),
\]
where the abbreviation $\vecvar{x}^{2}\Define\sum_{j=1}^{N}x_{j}^{2}$ is used.
In terms of the number $n_{j}\Define a_{j}^{\dagger}a_{j}$
creation $a_{j}^{\dagger}\Define\frac{1}{2{\rm i}\omega}
\bigl(p_{j}+{\rm i}\omega x_{j}\bigr)$ and annihilation
$a_{j}\Define{\rm i}\bigl(p_{j}-{\rm i}\omega x_{j}\bigr)$ operators,
the above relation is expressed by
\[
  {\rm e}^{-\frac{1}{2}\omega\vecvar{x}^{2}}
  \omega\bigl({\cal O}_{\rm E}-\frac{1}{2\omega}\nabla^{2}\bigr)
  {\rm e}^{\frac{1}{2}\omega\vecvar{x}^{2}}
  = \omega\sum_{j=1}^{N}n_{j}.
\]
In summary, successive applications of the similarity transformations
transform the Calogero Hamiltonian $\hat{H}_{\rm C}$
\myref{eqn:Calogero_Hamiltonian}
to the Hamiltonian of the decoupled $N$ quantum harmonic
oscillators.

Since the number operators, $n_{j}$, $j=1,\cdots,N,$ mutually commute,
these conserved operators are simultaneously diagonalizable. Thus the
non-degenerate simultaneous eigenfunctions of the number operators,
which are nothing but the number states,
form the non-symmetric orthogonal basis
of the Hilbert (Fock) space for the $N$ decoupled harmonic oscillators.
The similarity transformation indicates that
the non-symmetric orthogonal basis for the Calogero model can be constructed
exactly in the same manner.
Defining the creation, annihilation and number operators by
\begin{eqnarray*}
  & & \Cmae\;\;\; \hat{b}_{j}^{+} \Define 
  \phi_{\rm g}{\rm e}^{-\frac{1}{4\omega}{\cal O}_{\rm L}}
  {\rm e}^{\frac{1}{4\omega}\nabla^{2}}
  {\rm e}^{\frac{1}{2}\omega\vecvar{x}^{2}}a_{j}^{\dagger}
  {\rm e}^{-\frac{1}{2}\omega\vecvar{x}^{2}}
  {\rm e}^{-\frac{1}{4\omega}\nabla^{2}}
  {\rm e}^{\frac{1}{4\omega}{\cal O}_{\rm L}}\phi_{\rm g}^{-1},\\
  & & \Cmae\;\;\; \hat{b}_{j} \Define 
  \phi_{\rm g}{\rm e}^{-\frac{1}{4\omega}{\cal O}_{\rm L}}
  {\rm e}^{\frac{1}{4\omega}\nabla^{2}}
  {\rm e}^{\frac{1}{2}\omega\vecvar{x}^{2}}a_{j}
  {\rm e}^{-\frac{1}{2}\omega\vecvar{x}^{2}}
  {\rm e}^{-\frac{1}{4\omega}\nabla^{2}}
  {\rm e}^{\frac{1}{4\omega}{\cal O}_{\rm L}}\phi_{\rm g}^{-1},\\
  & & \Cmae\;\;\; \hat{n}_{j}^{\rm C}\Define
  \hat{b}_{j}^{\dagger}\hat{b}_{j},
\end{eqnarray*}
we can algebraically construct the non-symmetric simultaneous
eigenfunctions of all the number
operators $\hat{n}_{j}^{\rm C}$ as
\begin{equation}
  \Cmae\;\;\; |\lambda_{\sigma}\rangle = 
  \prod_{j=1}^{N}(\hat{b}_{j}^{+})^{\lambda_{\sigma(j)}}
  |0\rangle
  \Define M_{\lambda_{\sigma}}(\vecvar{x};1/a,\omega)\phi_{\rm g},\;\;
  |0\rangle \Define \phi_{\rm g},
  \label{eqn:Non-Symmetric_Orthogonal_Basis}
\end{equation}
where $\lambda$ and $\sigma$ are respectively
the Young diagram, $\lambda\Define\{
\lambda_{1}\geq\lambda_{2}\geq\cdots\geq\lambda_{N}\geq 0\}$, where
$\lambda_{k}$, $k=1,\cdots,N$, are integers, and the permutation,
$\sigma \in S_{N}$.
The function $M_{\lambda_{\sigma}}(\vecvar{x};1/a,\omega)$ is a
non-symmetric polynomial.
These non-symmetric functions are apparently non-degenerate simultaneous
eigenfunctions of the commuting number operators $\hat{n}_{j}^{\rm C}$.
Defining the bras in exactly the same way as we do
for the harmonic oscillators,
\begin{equation}
  \langle\lambda_{\sigma}| = \langle 0|
  \prod_{j=1}^{N}(\hat{b}_{j})^{\lambda_{\sigma(j)}},\;\;
  \langle 0| \Define \phi_{\rm g},
  \label{eqn:Dual_Basis}
\end{equation}
which is nothing but the dual of the above kets,
we can readily confirm the bras and the kets form the orthogonal basis,
\[
  \langle\lambda_{\sigma}|\mu_{\tau}\rangle =
  \delta_{\lambda_{\sigma},\mu_{\tau}}\prod_{j=1}^{N}\lambda_{j}!
  \langle 0|0\rangle .
\]
See Ref.\ \cite{Baker_2,vanDiejen_1} for the vacuum normalization.
Thus we have constructed the non-symmetric orthogonal
basis, or the Fock space,
for the Calogero model with distinguishable particles.

In order to compare the above basis with the known basis given by the
Hi-Jack polynomials~\cite{Ujino_1,Ujino_2,Ujino_3},
we rewrite the creation-annihilation operators
as
\begin{eqnarray*}
  & & x_{j} =
  {\rm e}^{\frac{1}{4\omega}\nabla^{2}}
  {\rm e}^{\frac{1}{2}\omega\vecvar{x}^{2}}a_{j}^{\dagger}
  {\rm e}^{-\frac{1}{2}\omega\vecvar{x}^{2}}
  {\rm e}^{-\frac{1}{4\omega}\nabla^{2}},\\
  & & \frac{\partial}{\partial x_{j}} =
  {\rm e}^{\frac{1}{4\omega}\nabla^{2}}
  {\rm e}^{\frac{1}{2}\omega\vecvar{x}^{2}}a_{j}
  {\rm e}^{-\frac{1}{2}\omega\vecvar{x}^{2}}
  {\rm e}^{-\frac{1}{4\omega}\nabla^{2}}.
\end{eqnarray*}
Then the polynomial parts of the kets
\myref{eqn:Non-Symmetric_Orthogonal_Basis} is rewritten as
\begin{equation}
  M_{\lambda_{\sigma}}(\vecvar{x};1/a,\omega)
  ={\rm e}^{-\frac{1}{4\omega}{\cal O}_{\rm L}}
  \vecvar{x}^{\lambda_{\sigma}},
  \label{eqn:Sogo_representation}
\end{equation}
where $\vecvar{x}^{\lambda_{\sigma}}\Define 
x_{1}^{\lambda_{\sigma(1)}}\cdots x_{N}^{\lambda_{\sigma(N)}}$.
A comment might be in order. In order to avoid the essential singularity,
we have to restrict the operand of the symmetric Lassalle operator
to the symmetric functions~\cite{Ujino_5}. On the other hand,
action of the non-symmetric Lassalle operator on a monomial yields
a polynomial with no essential singularity.
As was shown by Baker and Forrester~\cite{Baker_2},
the non-symmetric Hi-Jack polynomials~\cite{Baker_2,Kakei_1,Takemura_1}
is given by
\begin{equation}
  j_{\lambda_{\sigma}}(\vecvar{x};1/a,\omega)
  ={\rm e}^{-\frac{1}{4\omega}{\cal O}_{\rm L}}
  J_{\lambda_{\sigma}}(\vecvar{x};1/a),
  \label{eqn:Sogo-Lassalle_Formula}
\end{equation}
where $J_{\lambda_{\sigma}}(\vecvar{x};1/a)$ is the
non-symmetric Jack polynomial~\cite{Opdam,Bernard}.
The non-symmetric Jack polynomials have
triangular expansion forms with respect to the monomials,
\begin{equation}
  J_{\lambda_{\sigma}}(\vecvar{x};1/a) = \vecvar{x}^{\lambda_{\sigma}}
  +\Cmae\sum_{\mu_{\tau}<\lambda_{\sigma}}
  v_{\lambda_{\sigma},\mu_{\tau}}(a)\vecvar{x}^{\mu_{\tau}}
  \Define T_{\lambda_{\sigma},\mu_{\tau}}\vecvar{x}^{\mu_{\tau}}.
  \label{eqn:Jack}
\end{equation}
The order among the pairs of the Young diagrams and the permutations
$\lambda_{\sigma}$ are defined by
\[
  \mu_{\tau}<\lambda_{\sigma}\Leftrightarrow
  \left\{
    \begin{array}{ll}
      1) & \mu\stackrel{\rm D}{<}\lambda, \\
      2) & \mbox{when $\mu=\lambda$ then the first}\\
      & \mbox{non-vanishing difference}\\
      & \tau(i)-\sigma(i)<0,
    \end{array}
  \right.
\]
where the dominance order $\stackrel{\rm D}{<}$ is given as
\[
  \mu\stackrel{\rm D}{<}\lambda \Leftrightarrow \mu\neq\lambda,
  \; |\mu|=|\lambda|
  \mbox{ and }\sum_{k=1}^{l}\mu_{k}\leq\sum_{k=1}^{l}\lambda_{k},
\]
for all $l=1,\cdots,N$.
We have used here the symbol for the weight of the Young diagram,
$|\lambda|\Define\sum_{j=1}^{N}\lambda_{j}$.
Equations \myref{eqn:Sogo_representation} --
\myref{eqn:Jack} show that the non-symmetric
Hi-Jack polynomials and 
the new non-symmetric orthogonal basis are different, though
the latter also can be regarded as a multivariable generalization
of the Hermite polynomial with one additional parameter.
From the above expansion \myref{eqn:Jack}, we readily confirm that
the non-symmetric Jack polynomials are not the simultaneous eigenfunctions
of the operators $x_{j}\frac{\partial}{\partial x_{j}}$, $j=1,\cdots,N$,
and hence
the non-symmetric Hi-Jack polynomials are not the simultaneous eigenfunctions
of the commuting number operators $\hat{n}_{j}^{\rm C}$, $j=1,\cdots,N$.

It seems rather strange at first sight that
the two orthogonal bases are related by a not unitary but triangular matrix
$T_{\lambda_{\sigma},\mu_{\tau}}$. It reflects the fact that the
commuting number operators $\hat{n}_{j}^{\rm C}$ are not Hermitian
but self-dual with respect to the exchange of the
creation and annihilation operators
$\hat{b}_{j}^{+}\leftrightarrow\hat{b}_{j}$, $j=1,\cdots,N$.
That is why the bra $\langle \lambda_{\sigma}|$ is not given by the
corresponding ket itself, but should be identified with a ``rotation''
of the ket, $\sum_{\mu_{\tau},\nu_{\upsilon}}T_{\lambda_{\sigma},\mu_{\tau}}
T_{\mu_{\tau},\nu_{\upsilon}}|\nu_{\upsilon}\rangle$ in the inner product.
On the other hand, 
the Dunkl operator formulation~\cite{Ujino_1,Ujino_2,Ujino_3,%
Dunkl_1,Polychronakos_1} yields 
the Hermitian commuting conserved operators for the Calogero model.
The Dunkl operators for the Calogero model are listed as
\begin{eqnarray*}
  & & \hat{\alpha}_{l}^{\dagger}\Define-\frac{\rm i}{2\omega}\Bigl(
  p_{l}+{\rm i}a\sum_{\stackrel{\scriptstyle k=1}{k\neq l}}^{N}
  \frac{1}{x_{l}-x_{k}}K_{lk}+{\rm i}\omega x_{l}\Bigr),\\
  & & \hat{\alpha}_{l}\Define{\rm i}\Bigl(
  p_{l}+{\rm i}a\sum_{\stackrel{\scriptstyle k=1}{k\neq l}}^{N}
  \frac{1}{x_{l}-x_{k}}K_{lk} -{\rm i}\omega x_{l}\Bigr),\\
  & & \hat{d}_{l}\Define\hat{\alpha}_{l}^{\dagger}\hat{\alpha}_{l}
  +a\sum_{j=1}^{l-1}(K_{jl}-1),\;\;\; [\hat{d}_{l},\hat{d}_{m}]=0.
\end{eqnarray*}
In terms of the Dunkl operators,
the Hamiltonian for the Calogero model \myref{eqn:Calogero_Hamiltonian}
is given by $\hat{H}_{\rm C}-E_{\rm g}=\omega \hat{I}_{1}$, where
$\hat{I}_{n}\Define\sum_{l=1}^{N}(\hat{d}_{l})^{n},\;\; n=1,\cdots,N$.
And the Hermitian commuting conserved operators are given by the
$\hat{d}_{l}$-operators (and also $\hat{I}_{n}$).
The polynomial parts of the non-symmetric simultaneous eigenfunctions
for all the above commuting Dunkl operators $\hat{d}_{l}$,
or equivalently, the conserved operators $\hat{I}_{n}$, are uniquely
identified as the non-symmetric Hi-Jack polynomials, which give an orthogonal
basis of the Calogero model with respect to the conventional Hermitian inner
product. The non-symmetric simultaneous eigenfunctions 
\myref{eqn:Non-Symmetric_Orthogonal_Basis} are not the simultaneous
eigenfunctions for these commuting conserved operators.

{} From the discussions above, we conclude that the Calogero Hamiltonian
has two sets of commuting conserved operators which are algebraically
inequivalent to each other.  We also conclude that two different
conserved operators respectively picked up from the two different sets
do not commute $[\hat{n}_{j}^{\rm C},\hat{I}_{k}]\neq 0$,
for $k\neq 1$, or equivalently,
$[\hat{n}_{j}^{\rm C},\hat{d}_{k}]\neq 0$.
The Hilbert space of the Calogero Hamiltonian also has two different
orthogonal bases that respectively correspond to the simultaneous
eigenfunctions for the two sets of commuting conserved operators.
This peculiar fact must be due to the large degeneracy of the
eigenvalue of the Calogero Hamiltonian \myref{eqn:Calogero_Hamiltonian},
\[
  \hat{H}_{\rm C}|\lambda_{\sigma}\rangle
  = \bigl(\omega|\lambda|+E_{\rm g}\bigr)|\lambda_{\sigma}\rangle.
\]
For a particular eigenvalue, say $\omega n +E_{\rm g}$,
the degeneracy is given by
the number of pairs of Young diagrams and permutations
$\lambda_{\sigma}$ such that $|\lambda|=n$.
Existence of two inequivalent sets of conserved operators and two
different simultaneous eigenfunctions implies  
some hidden dynamical symmetry of the
Calogero model, as is the case with the hydrogen atom
that has the $O(4)$ dynamical symmetry related to the angular
momentum and the Runge-Lenz-Pauli vector~\cite{Schiff_1}.

We should note that our discussions so far on the ($A_{N-1}$-)Calogero model
also holds for the $B_{N}$-Calogero model~\cite{Yamamoto_1},
\begin{eqnarray*}
  \hat{H}_{B_{N}} & = & \frac{1}{2}\sum_{j=1}^{N}
  \bigl(p_{j}^{2}+\omega^{2}x_{j}^{2}\bigr)
  +\frac{1}{2}\sum_{j=1}^{N}\frac{(b^{2}-bt_{j})}{x_{j}^{2}}\\
  & & +\frac{1}{2}\sum_{\stackrel{\scriptstyle j,k=1}{j\neq k}}^{N}
  \Bigl(\frac{a^{2}-aK_{jk}}{(x_{j}-x_{k})^{2}}
  +\frac{a^{2}-at_{j}t_{k}K_{jk}}{(x_{j}+x_{k})^{2}}\Bigr).
\end{eqnarray*}
where the reflection operator $t_{j}$ is defined by
$(t_{j}f)(\cdots,x_{j},\cdots)=f(\cdots,-x_{j},\cdots)$.
As is similar to the way of Gurappa and Panigrahi for the symmetric case%
~\cite{Gurappa_2},
we can decompose the above Hamiltonian into that of the decoupled quantum
particles.
By the following similarity transformation, we have
\[
  \phi_{\rm g}^{-1}(\hat{H}_{B_{N}}-E_{\rm g}^{B_{N}})\phi_{\rm g}
  =\omega{\cal O}_{\rm E}
  -\frac{1}{2}{\cal O}_{\rm L}^{B_{N}},
\]
where the ground state, the ground state energy and the non-symmetric
$B_{N}$-Lassalle operator are~\cite{Baker_2}
\begin{eqnarray*}
  & & \phi_{\rm g}(\vecvar{x})
  =\prod_{1\leq j<k\leq N}|x_{j}^{2}-x_{k}^{2}|^{a}
  \prod_{j=1}^{N}|x_{j}|^{b}\exp\Bigl(-\frac{1}{2}\omega
  \sum_{j=1}^{N}x_{j}^{2}\Bigr),\\
  & & E_{\rm g}^{B_{N}}=\frac{1}{2}N\omega(1+2a(N-1)+2b),\\
  & & {\cal O}_{L}^{B_{N}}=
  \sum_{j=1}^{N}\Bigl(\frac{\partial^{2}}{\partial x_{j}^{2}}
  +\frac{2b}{x_{j}}\frac{\partial}{\partial x_{j}}
  +\frac{b}{x_{j}^{2}}(t_{j}-1)\Bigr)\\
  & & \Usr\Cusr +a\sum_{\stackrel{\scriptstyle j,k=1}{j\neq k}}^{N}
  \Bigl(
  \frac{2}{x_{j}^{2}-x_{k}^{2}}
  \bigl(x_{j}\frac{\partial}{\partial x_{j}}
  -x_{k}\frac{\partial}{\partial x_{k}}
  \bigr)\nonumber\\
  & & \Usr\Cusr +\frac{K_{jk}-1}{(x_{j}-x_{k})^{2}}
  +\frac{t_{j}t_{k}K_{jk}-1}{(x_{j}+x_{k})^{2}}\Bigr).
\end{eqnarray*}
Since the Hamiltonian is transformed to the Euler operator by
\[
  {\rm e}^{\frac{1}{4\omega}{\cal O}_{\rm L}^{B_{N}}}
  \bigl(\omega{\cal O}_{\rm E}
  -\frac{1}{2}{\cal O}_{\rm L}^{B_{N}}\bigr)
  {\rm e}^{-\frac{1}{4\omega}{\cal O}_{\rm L}^{B_{N}}}
  =\omega{\cal O}_{\rm E},
\]
the polynomial part of the non-symmetric orthogonal basis
for the $B_{N}$-Calogero Hamiltonian $\hat{H}_{B_{N}}$ is given by
\[
  Y_{\lambda_{\sigma}}(\vecvar{x};1/a,1/b,\omega)
  ={\rm e}^{-\frac{1}{4\omega}{\cal O}_{\rm L}^{B_{N}}}
  \vecvar{x}^{\lambda_{\sigma}},
\]
in a similar way to Eq. \myref{eqn:Sogo_representation}.
They are  the simultaneous eigenfunctions of the operators
$x_{j}\frac{\partial}{\partial x_{j}}$.
Also in a similar way to the Calogero model,
the non-symmetric orthogonal basis is shown to be different from
the known orthogonal basis, namely, the non-symmetric
generalized Laguerre polynomials~\cite{Baker_2,Kakei_2}.
And the number operators are algebraically inequivalent to the known
commuting conserved operators constructed by Dunkl operator formulation.

So far, we have not found a similar transformation of the Sutherland model
into a decoupled system. This seems rather strange
at first because the commuting
conserved operators and the Dunkl operators for the
Calogero and Sutherland models are
known to share the same algebraic structure,
and become exactly the same in the limit,
$\omega\rightarrow\infty$~\cite{Ujino_2}
(strictly speaking, we have to remove the action to the ground state
when we consider the correspondence).
The difference of the two models is the structure of the Hamiltonian.
While the Calogero Hamiltonian is the simplest
conserved operator $\hat{I}_{1}$, the Sutherland Hamiltonian corresponds
to the second conserved operator $\hat{I}_{2}$. We have proved that the
second conserved operator $\hat{I}_{2}$ can not be constructed from the
number operators $\hat{n}_{j}^{\rm C}$.
We think that the point causes the essential
difficulty in the application of such a similarity transformation method
to the Sutherland model.

Let us summarize this letter. We have shown an algebraic construction
of the non-symmetric orthogonal basis for the Calogero model in completely
parallel way to that of the quantum harmonic oscillators. The basis
consists of the simultaneous eigenfunctions of the commuting number
operators, which is different from the known orthogonal basis
spanned by the non-symmetric Hi-Jack polynomials. The commuting
number operators are algebraically inequivalent to the known commuting
conserved operators given by the Dunkl operator formulation. Existence
of the two different sets of commuting conserved operators and the two
different orthogonal bases implies
some hidden dynamical symmetry behind the model.
We have also shown that the same
story holds for the $B_{N}$-Calogero model. We have discussed on
the difficulty in the application of this similarity-transformation method
to the Sutherland model. We hope that the new
non-symmetric orthogonal bases will shed new
light on the study of the eigenfunctions and the correlation functions
of the models with or without spins.

The authors appreciate encouraging communications with K.~Sogo and
F.~G\"{o}hmann.
One of the authors (HU) appreciates the Research Fellowships of the
Japan Society for the Promotion of Science for Young Scientists.


\end{document}